\documentclass[iop]{emulateapj}

\newcommand{\sgr}{SGR~0501+4516~}

\newcommand{\myemail}{physicsmyl@gmail.com}

\shorttitle{X-RAY OBSERVATIONS OF SGR~0501+4516}
\shortauthors{Mong \& Ng}
\submitted{ApJ, in press}

\begin{document}


\title{X-Ray Observations of Magnetar \sgr from Outburst to Quiescence}


\author{Y.-L.\,Mong and C.-Y.\,Ng}
\affil{Department of Physics, The University of Hong Kong, Pokfulam
Road, Hong Kong;\\ 
\myemail, ncy@astro.physics.hku.hk}



\begin{abstract}
Magnetars are neutron stars having extreme magnetic field strengths.
Study of their emission properties in quiescent state can help understand
effects of a strong magnetic field on neutron stars.
\sgr is a
magnetar that was discovered in 2008 during an outburst, which has recently returned to
quiescence. We report
its spectral and timing properties measured with new and archival observations
from
the \emph{Chandra X-Ray Observatory}, \emph{XMM-Newton}, and \emph{Suzaku}.
We found that the quiescent spectrum is best fit by a power-law plus two
blackbody model, with temperatures
of $kT_{\rm low}\sim0.26\,{\rm keV}$ and $kT_{\rm
high}\sim0.62\,{\rm keV}$.
We interpret these two blackbody components as emission from a hotspot and the
entire surface.
The hotspot radius shrunk from $1.4\,{\rm km}$ to
$0.49\,{\rm km}$ since the
outburst, and there was a significant correlation between its area and the X-ray
luminosity,
which agrees well with the prediction by the twisted magnetosphere model.
We applied the two-temperature spectral model to all magnetars in quiescence
and
found that it could be a common feature among the population.
Moreover, the temperature of the cooler blackbody shows a
general trend with the magnetar field strength, which supports the simple
scenario of heating by magnetic field decay.
\end{abstract}





\section{\bf{Introduction}} \label{sec:intro}
Magnetars are non-accreting neutron stars with long spin periods ($P\sim2$--$12\,{\rm
s}$) and the largest spin-down rates ($\dot{P}\sim10^{-13}$--$10^{-10}\,{\rm
s}\,{\rm
s}^{-1}$) among the pulsar population. Most of them have spin-down inferred magnetic field
strength, $B$,
up to $\sim10^{15}\,{\rm G}$. It is generally believed that
magnetars are young neutron stars and some are found inside Supernova Remnants
(SNRs).
Magnetars usually have persistent X-ray luminosity,
$L_{\rm X}\sim10^{34-36}\,{\rm erg}\,{\rm s}^{-1}$, much larger than
their rotational energy loss rate $\dot{E}$,
and they occasionally exhibits violent bursting activities \citep[see review
by][]{kb17}.
In order to explain the properties of this pulsar class, magnetar models have
been developed. The most popular one is the twisted
magnetosphere model \citep{td95,td01,bel09,bel11}.
It suggests that the toroidal magnetic field could exist in the stellar
crust.
If the internal magnetic field is strong enough,
it could tear the crust followed by twisting the
crust-anchored external field \citep{td95,tdw00,tyk02}.
In addition, a starquake arising from the plastic deformation of the crust would
cause magnetar
bursts due to magnetic reconnection \citep{td95,pbh12,pbh13}.
Persistent X-ray emission of magnetars could be explained by the magnetic field decay
\citep{tyk02,plm07}. Meanwhile, the magneto-thermal evolution theory suggests
that the field decay could be
enhanced due to the changes in the conductivity and the magnetic diffusivity of magnetars
\citep{vrp13}. As a consequence, magnetars are observed to have higher surface
temperature and X-ray luminosity than canonical pulsars.
In general, soft X-ray spectra of magnetars can be
described by an absorbed blackbody model with temperature
$kT\sim0.3$--$0.6\,{\rm keV}$ plus an
additional power-law with photon index $\Gamma\sim2$--$4$ or
another blackbody component with $kT\sim0.7\,{\rm keV}$
\citep[see][]{ok14,kb17}. It indicates that the
soft X-ray emission could be contributed by thermal emission and some
non-thermal radiation processes, such as synchrotron or inverse-Compton
scattering.\\
\indent \object{SGR~0501+4516} is a magnetar discovered with the Burst Alert
Telescope (BAT) on board \emph{Swift} on 2008\,August\,22 due to
a series of short bursts \citep{bbb08}. X-ray pulsations were detected with a
period of $P\sim5.7\,{\rm s}$ \citep{rit09}.
After the discovery, the source was subsequently identified in an archival
\emph{ROSAT} observation taken in 1992.
The soft X-ray flux was $\sim80$ times higher in the outburst when compared to
the 1992 observation \citep{rit09}. The hard X-ray tail above $10\,{\rm keV}$
was first discovered with \emph{INTEGARL} right after the outburst \citep{rit09}. It had also been
detected with \emph{Suzaku} observation \citep{ern10}.
From the spin period and spin-down rate, $B$
was estimated to be $2\times10^{14}\,{\rm G}$ \citep{wgk08}. The soft X-ray
spectrum of \sgr
below $10\,{\rm keV}$ could be described by an absorbed blackbody model with a
power-law component, using
\emph{XMM-Newton} observations obtained in the first year after the outburst
\citep{rit09,cpr14}.
The X-ray spectral properties
from 2008 to 2013 were also measured with four \emph{Suzaku} observations
\citep{esk17}, but it is interesting to note that
the results are different from those reported in other literature, including
a smaller hydrogen column density,
lower blackbody temperature, larger radius, and softer power-law photon index
\citep{rit09,gwk10,cpr14}.\\
\begin{deluxetable*}{ccccc}
\tablecaption{Observations of SGR 0501+4516 Used in Our Analysis.
\label{tab:obs}}
\tablehead{\colhead{Date} & \colhead{Observatory
(Instruments)} & \colhead{ObsID} &
\colhead{Mode} & \colhead{Net Exposure (ks)}
}
\startdata
2008 Aug 31 & \emph{XMM-Newton} (PN) &
0552971201 &
SW & $10.2$ \\
2008 Sep 02 & \emph{XMM-Newton} (PN) &
0552971301 &
SW & $20.5$ \\
2008 Sep 25 & \emph{CXO} (HRC-I) &
9131 &
-- & $10.1$ \\
2008 Sep 30 & \emph{XMM-Newton} (PN/MOS1/MOS2) &
0552971401 &
LW/SW/SW &
$30.1/32.3/32.3$ \\
2009 Aug 30 & \emph{XMM-Newton} (PN/MOS1/MOS2) &
0604220101 &
SW/FF/SW &
$53.9/52.4/53.1$ \\
2012 Dec 09 & \emph{CXO} (ACIS-S\tablenotemark{a})&
15564 &
TE & $14.0$ \\
2013 Apr 03 & \emph{CXO} (ACIS-S\tablenotemark{a})&
14811 &
TE & $13.7$ \\
2013 Aug 31 & \emph{Suzaku} (XIS0/XIS1/XIS3) & 408013010
& Normal & $36.0/41.1/41.2$
\enddata
\tablecomments{
\tablenotetext{a}{Made in the sub-array mode with only one-eighth of CCD\,7.}}
\end{deluxetable*}
\indent Until now, there is no accurate distance measurement for SGR~0501+4516.
As magnetars are young pulsars, \sgr is expected to be located
close to the
spiral arm of the Galaxy.
The line of sight intercepts the Perseus and Outer arms of the Galaxy, at
distances of $\sim2.5$ and $\sim5\,{\rm kpc}$,
respectively.
In this paper, we assume the distance $d=5\,{\rm kpc}$. In addition, there
exists a supernova remnant
(SNR) G160.9+2.6, $\sim80'$ north of \sgr \citep{gc08,gwk10}. The
distance and age of the SNR were estimated as
$800\pm400\,{\rm pc}$ and $4000$--$7000\,{\rm years}$ \citep{lt07}.
\citet{gwk10} proposed that \sgr could be associated
with G160.9+2.6. Leaving the distance aside, if
this is the case, the magentar should have a large proper motion of
$0\farcs7$--$1\farcs2\,{\rm yr}^{-1}$ to the south. \\
\indent In this paper, we used new X-ray observations to show that
SGR 0501+4516 had returned to quiescence in 2013,
five years after the outburst, and we report on its spectral and timing
properties during flux relaxation.
We also analyzed
archival observations to investigate the
long-term evolution.
\section{\bf{Observations and data reduction}} \label{sec:obs}
There are eight X-ray observations used in this study (see
Table\,\ref{tab:obs}). We obtained two new
observations obtained with the Advanced CCD Imaging Spectrometer (ACIS)
on board \emph{Chandra X-ray Observatory (CXO)} on 2012\,December\,9 and
2013\,April\,3. Both of them were made in the Time Exposure (TE) mode for
14\,ks using only one-eighth of the CCD, providing a fast frame time of
0.4\,s. This allows us to obtain a crude pulse
profile for this $\sim5.67\,{\rm s}$ period pulsar.
By inspecting the light curves, no bursts from the source or background flares were detected during
the exposures. We checked that pile-up was negligible
in both observations. In addition to these two ACIS observations, a
\emph{Chandra}
High Resolution Camera (HRC) observation taken on 2008\,September\,25 was also used
to measure the source position only.
All \emph{Chandra} data were reprocessed with \texttt{chandra\_repro}
in CIAO 4.8 with CALDB 4.7.4 before performing any analysis.\\
\indent There were six \emph{XMM-Newton} observations after the discovery of
the source. We only analyzed the latest
four from 2008\,August\,31 to 2009\,August\,30 because
SGR 0501+4516 showed strong bursting activities during the two earliest
observations.
The source was still bright 11 days after the outburst; the pile-up effect
was an issue in the
MOS data obtained on 2008\,August\,31 and September\,2 and hence only the PN
data were
used in these two observations.
We first reprocessed all the data by the tasks
\texttt{epchain}/\texttt{emchain}
in XMMSAS version 1.2. In the analysis,
only PATTERN\,$\leq4$ events of the PN data and PATTERN\,$\leq12$ events in
the MOS data were used.
We also used the standard screening for
the MOS (FLAGS~=~\#XMMEA\_EM) and
PN (FLAGS~=~\#XMMEA\_EP) data.
After removal of periods with
background flares, we obtained net exposures ranging from $10.2$ to
$53.9\,{\rm ks}$ (see Table\,\ref{tab:obs}).\\
\indent We also used the latest \emph{Suzaku} data in the archive taken on
2013\,August\,31,
to combine with the \emph{Chandra} data to better constrain the quiescent
spectral properties.
In order to focus on the soft X-ray spectral properties,
only the data obtained with the XIS were used (see Table\,\ref{tab:obs}).
The XIS data were reprocessed using \texttt{xisrepro} in HEAsoft 6.20
with standard screening criteria. We inspected the light curves
to verify that no bursts were detected throughout the observation with
$\sim40\,{\rm ks}$.
\section{\bf{ANALYSIS AND RESULTS}}
\subsection{Imaging and Astrometry}
We measured the position of \sgr in all
\emph{Chandra} data using the
task \texttt{celldetect} and obtained a consistent result of
$\alpha$=5:01:06.8,
$\delta$=+45:16:34 (J2000) within the uncertainty.
The measurement uncertainties in the $90\%$ confidence level have radii $0\farcs4$ (HRC) and $0\farcs5$ (ACIS).
As the ACIS images were taken in the sub-array mode with a small field of
view, we did not find any background sources to align the two images.
Therefore, we also need to consider the absolute astrometric accuracy of
\emph{Chandra}, which is $0\farcs8$ at the $90\%$ confidence
level\footnote{\url{http://cxc.harvard.edu/cal/ASPECT/celmon/}}.
This gives an upper limit of the proper motion of
$0\farcs32\,{\rm yr}^{-1}$ ($90\%$ confidence level),
rejecting the suggestion that SGR 0501+4516 was born at the center of SNR
G160.9+2.6 \citep{gwk10}.\\
\indent Finally, we simulated a model point spread function for ACIS data with
ChaRT\footnote{\url{http://cxc.harvard.edu/ciao/PSFs/chart2/}} using the
best-fit spectrum (see Section\,\ref{sec:spe_ana} below)
and confirmed that the radial profile is fully consistent with that of the
real data, indicating no extended emission was found near the magnetar.
\subsection{Timing Analysis}\label{sec:tim_ana}
\begin{figure}[ht!]
\centering
\includegraphics[width=0.47\textwidth]{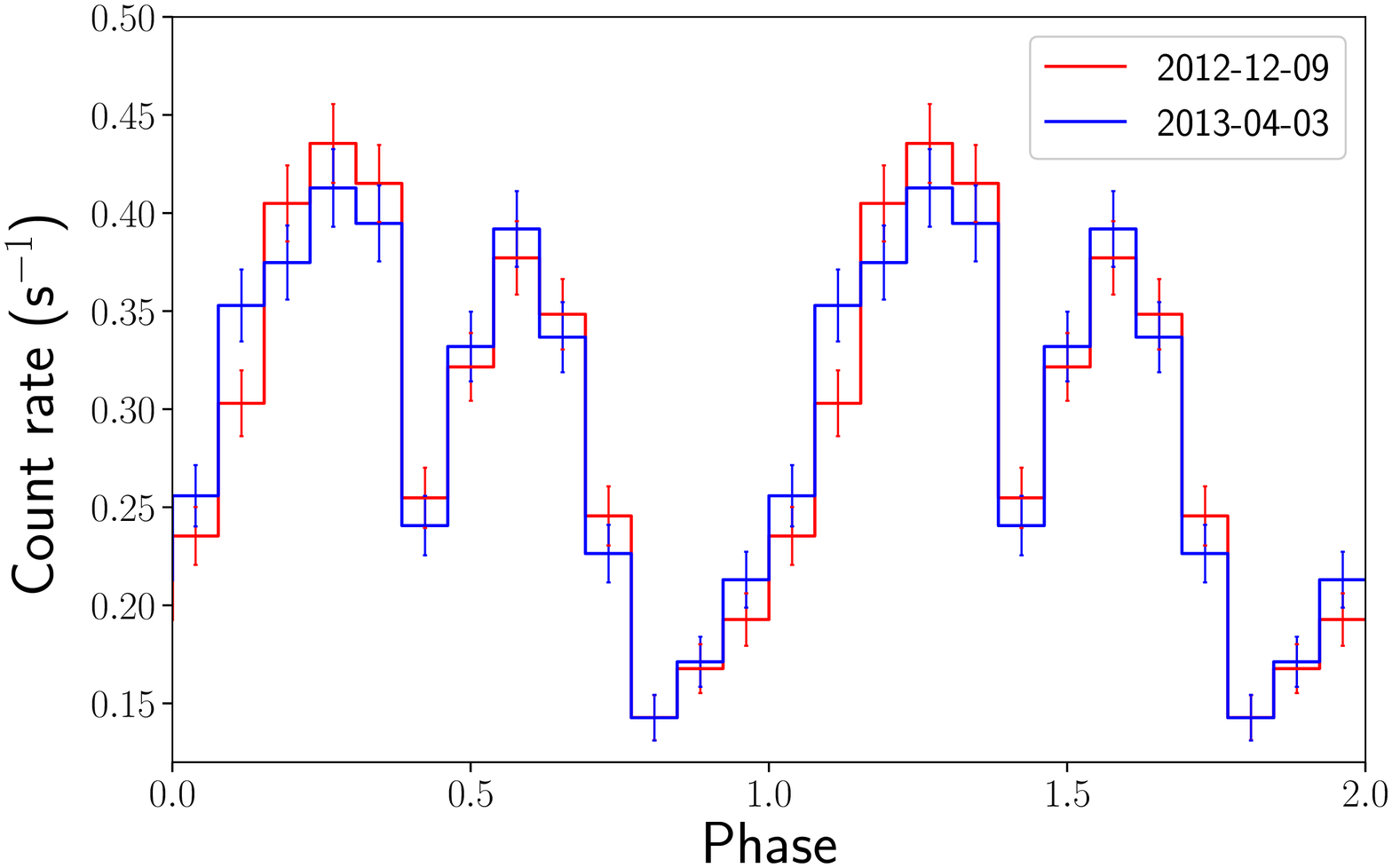}
\caption{Pulse profiles of SGR 0501+4516 in the energy range of
$0.5$--$7\,{\rm keV}$ for the latest \emph{Chandra} observations.
The two profiles were aligned manually by matching
the brightest bin. The uncertainties are at
the $1\sigma$ level.
}\label{fig:pulse_profile}
\end{figure}
\begin{figure}[ht!]
\centering
\includegraphics[width=0.45\textwidth]{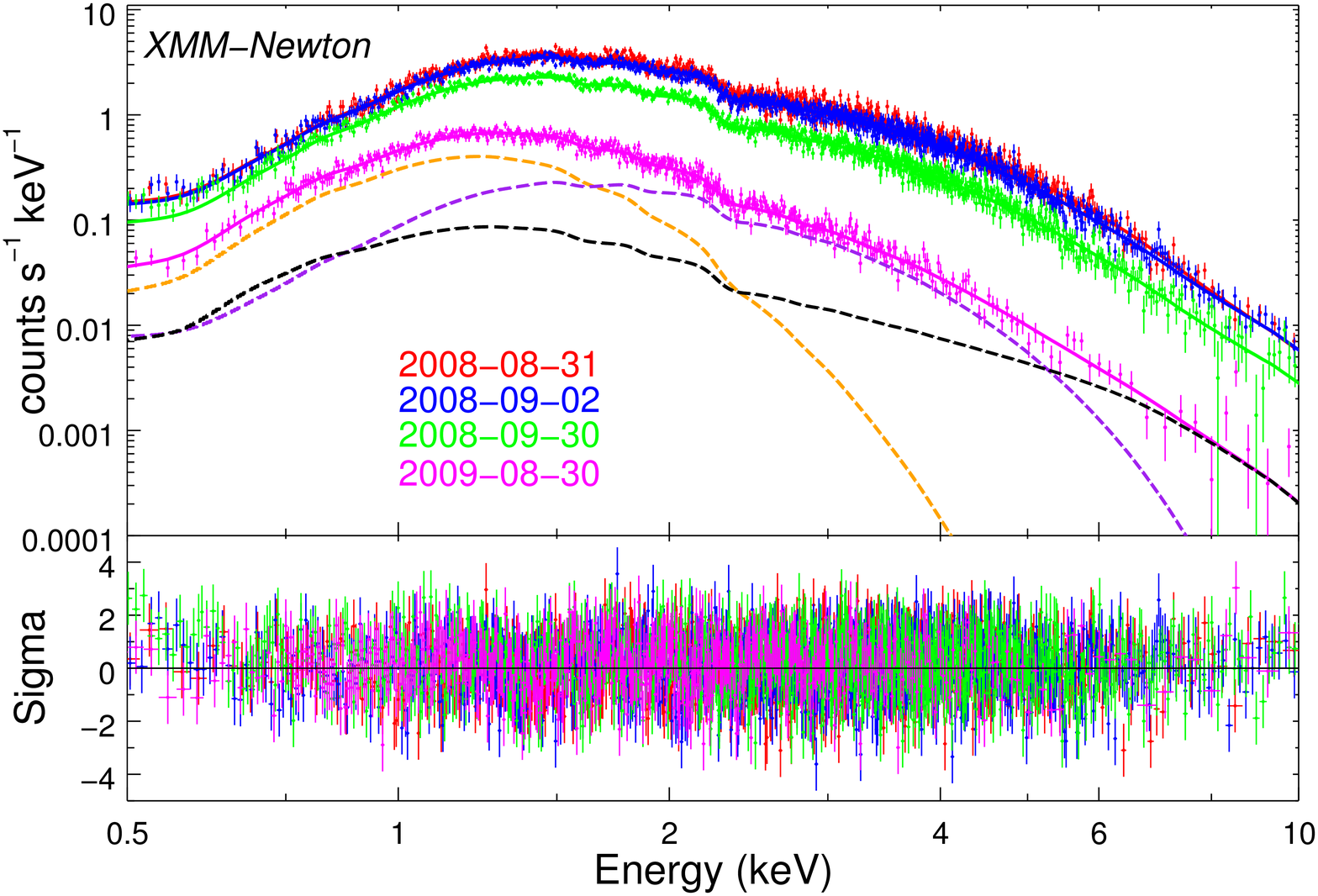}
\caption{\emph{XMM-Newton} PN spectra of SGR 0501+4516. The solid lines
indicate the
best-fit 2BB+PL model on different epochs. The orange, purple, and black
dashed
lines indicate the low temperature BB, high temperature BB, and
PL components of the 2009 August 30 spectrum, respectively.}\label{fig:xmm_fits}
\includegraphics[width=0.45\textwidth]{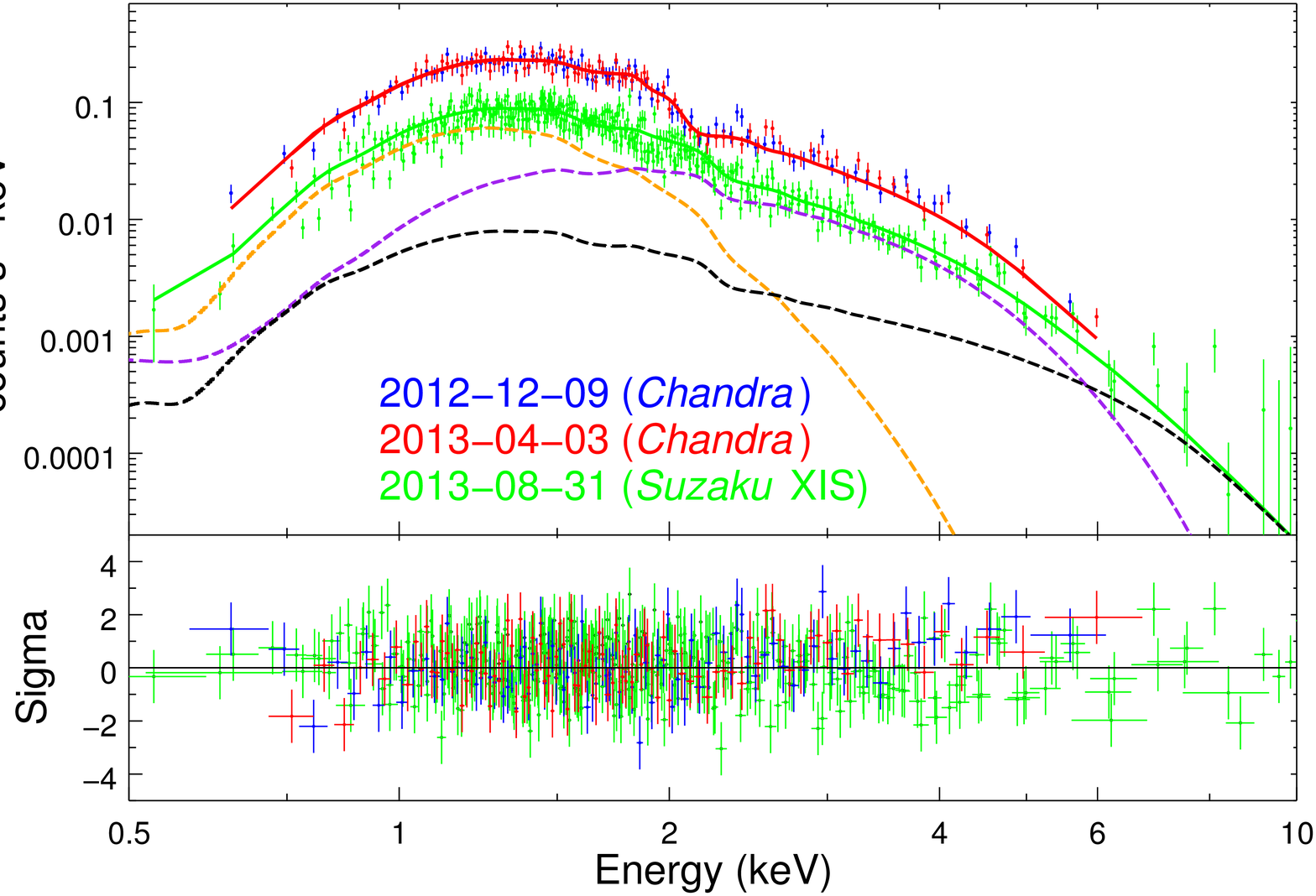}
\caption{\emph{Chandra} and \emph{Suzaku} spectra of SGR 0501+4516.  All
solid lines indicate the same best-fit 2BB+PL model. The different shape is
due to different responses of the
instruments. The orange, purple, and black dashed lines indicate the low
temperature BB, high
temperature BB, and PL components, respectively, with the \emph{Suzaku} XIS
response.}\label{fig:suzaku_chandra}
\end{figure}
We extracted the source photons from the two new \emph{Chandra} observations
by using a $2\farcs5$ radius aperture and obtained 4149 and 4043 counts,
respectively,
in the $0.5$--$7\,{\rm keV}$ energy range. The estimated
background photon counts in the source region are $\sim0.6$ for both
observations.
We then applied a barycentric
correction to the photon arrival times.
We employed the $\chi^2$-test after epoch folding
\citep{lea87} and found periods of
$P=5.76286(8)\,{\rm s}$ and $P=5.76299(9)\,{\rm s}$ for 2012\,December\,9 and
2013\,April\,3 data, respectively.
The $1\sigma$ uncertainties quoted here were estimated using the simulation results from
\citet{lea87}.
We used the best-fit periods to generate the pulse profiles for
both \emph{Chandra} observations. As the frame time of our
observations was $0.4\,{\rm s}$, we only divided the pulse period of
$P=5.76\,{\rm s}$ into 13 phase bins.
Figure\,\ref{fig:pulse_profile} shows the pulse profile, which has a
double-peaked shape. The pulse   
profile between the two observations did not show any obvious variations,
suggesting that
the source had already returned to quiescence
in 2013.\\
\indent As the dates of the two new \emph{Chandra}
observations were separated by too far apart, we were unable to measure the
spin-down rate $\dot{P}$ by phase coherent timing analysis. Meanwhile, the
uncertainties of individual timing measurements
were too large so that $\dot{P}$ could not be obtained from our \emph{Chandra}
observations.
We found that the two periods measured in 2012 and 2013 are formally
consistent with each other
after accounting for the uncertainties; however, they are different from the value obtained in the 2009 observation
\citep{cpr14}.
Comparing our results with the spin period $P=5.7622571(2)$ measured in 2009,
we obtained
$\dot{P}=6(1)\times10^{-12}\,{\rm s}\,{\rm s}^{-1}$ at the 
$1\sigma$ confidence level from 2009 to 2013, which is
compatible with
$5.94(2)\times10^{-12}\,{\rm s}\,{\rm s}^{-1}$ reported by \citet{cpr14}.
\subsection{Spectral Analysis}\label{sec:spe_ana}
We extracted the source spectrum from the \emph{Chandra} observations using
the same $2\farcs5$ radius apertures as in
the timing analysis above.
For the \emph{XMM-Newton} and \emph{Suzaku} XIS spectra, we used apertures of
$36''$ and $1\farcm8$ radius, respectively.
We chose a larger region far from the source on the same CCD as the background
region. We restricted
the analysis in the energy range of $0.5$--$10\,{\rm keV}$
for \emph{XMM-Newton} and \emph{Suzaku} data, and $0.5$--$7\,{\rm keV}$ for
\emph{Chandra} data to optimize the signal-to-noise ratio. We grouped the
spectra with a minimum of 30 counts
per energy bin.\\
\begin{turnpage}
\begin{deluxetable*}{lcccccccccc}[ht!]
\centering
\tablecaption{Best-Fit Spectral Parameters for SGR~0501+4516 with
Uncertainties
at the $90\%$
Confidence Level\label{tab:pas_results}}
\tablecolumns{10}
\tablehead{\colhead{Date} & \colhead{$N_{\rm H}$} &
\colhead{$kT_{\rm low}$} & \colhead{$R_{\rm low}$\tablenotemark{a}}
& \colhead{$F_{\rm low}$\tablenotemark{b} ($10^{-11}$}
& \colhead{$kT_{\rm high}$} & \colhead{$R_{\rm high}$\tablenotemark{a}} &
\colhead{$F_{\rm high}$\tablenotemark{b} ($10^{-11}$} & \colhead{$\Gamma$}
& \colhead{$F_{\rm PL}$\tablenotemark{b} ($10^{-11}$} & $\chi^2/{\rm dof}$ \\
 & \colhead{$(10^{22}\,{\rm cm}^{-2})$} & \colhead{(keV)} & \colhead{(km)} &
\colhead{erg\,cm$^{-2}\,{\rm s}^{-1}$)} & \colhead{(keV)} &
\colhead{(km)} & \colhead{${\rm erg}\,{\rm cm}^{-2}\,{\rm s}^{-1}$)} &
&  \colhead{erg\,cm$^{-2}\,{\rm s}^{-1}$)} &
}
\startdata
\cutinhead{BB+PL model}
2008 Aug 31\tablenotemark{c}& $1.34\pm0.06$ & \nodata & \nodata & \nodata &
$0.70\pm0.02$
& $1.45_{-0.08}^{+0.09}$ & $1.48\pm0.08$ & $2.9\pm0.1$ & $1.63\pm0.08$ & $764.9/741$\\
2008 Sep 02\tablenotemark{c} & $1.29_{-0.05}^{+0.04}$ & \nodata & \nodata & \nodata &
$0.68\pm0.01$ & $1.46\pm0.06$ & $1.34\pm0.05$ &
$2.85\pm0.07$ & $1.46\pm0.06$ & $947.0/915$ \\
2008 Sep 30\tablenotemark{d} & $1.36\pm0.03$ & \nodata & \nodata & \nodata &
$0.66\pm0.01$ & $1.03\pm0.04$ & $0.57\pm0.02$ & $3.15_{-0.05}^{+0.06}$
& $0.87\pm0.02$ & $2367.5/2169$ \\
2009 Aug 30\tablenotemark{d} & $1.43\pm0.04$ & \nodata & \nodata & \nodata &
$0.56\pm0.02$ & $0.56\pm0.05$ & $0.072_{-0.007}^{+0.008}$ &
$4.0\pm0.1$ & $0.21\pm0.01$ & $1308.3/1227$ \\
2013 Jun 23\tablenotemark{e} & $1.43_{-0.08}^{+0.09}$ & \nodata & \nodata & \nodata &
$0.63_{-0.05}^{+0.04}$ & $0.34_{-0.05}^{+0.06}$ & $0.05\pm0.01$ & $3.9_{-0.2}^{+0.3}$ & $0.15\pm0.01$
 & $626.9/603$ \\
\hline
\cutinhead{2BB+PL model}
2008 Aug 31\tablenotemark{c} & $0.90\pm0.02$\tablenotemark{f} & $0.35_{-0.02}^{+0.03}$ &
$4.6_{-0.5}^{+0.6}$ & $0.55\pm0.09$ & $0.75\pm0.02$ &
$1.4\pm0.1$ & $2.14_{-0.09}^{+0.07}$ & $1.33\tablenotemark{g}$ &
$0.42\pm0.07$ & $5911.5/5653$\tablenotemark{f} \\
2008 Sep 02\tablenotemark{c} & $0.90\pm0.02$\tablenotemark{f} & $0.31\pm0.02$ &
$5.2_{-0.6}^{+0.7}$ & $0.38\pm0.05$ & $0.71_{-0.02}^{+0.01}$ &
$1.56_{-0.06}^{+0.10}$ & $1.99_{-0.05}^{+0.04}$ &
$1.33\tablenotemark{g}$ & $0.45\pm0.04$ &
$5911.5/5653$\tablenotemark{f} \\
2008 Sep 30\tablenotemark{d} & $0.90\pm0.02$\tablenotemark{f} & $0.31\pm0.01$
& $4.8_{-0.2}^{+0.4}$ & $0.30\pm0.02$ & $0.69_{-0.01}^{+0.02}$ & $1.13_{-0.06}^{+0.05}$ &
$0.95\pm0.02$ &
$1.33\tablenotemark{g}$ & $0.20\pm0.02$ & $5911.5/5653$\tablenotemark{f} \\
2009 Aug 30\tablenotemark{d} & $0.90\pm0.02$\tablenotemark{f} & $0.25\pm0.02$ &
$4.4_{-0.4}^{+0.8}$ & $0.085\pm0.008$ & $0.56_{-0.02}^{+0.06}$ & $0.7\pm0.1$
& $0.14\pm0.01$ & $2.6_{-2.5}^{+0.4}$ & $0.06_{-0.02}^{+0.01}$ &
$5911.5/5653$\tablenotemark{f} \\
2013 Jun 23\tablenotemark{e} & $0.90\pm0.02$\tablenotemark{f} & $0.26_{-0.02}^{+0.01}$ &
$3.7_{-0.7}^{+0.3}$ & $0.07\pm0.01$ & $0.62_{-0.04}^{+0.03}$ &
$0.49_{-0.10}^{+0.05}$ &
$0.10_{-0.02}^{+0.01}$ &
$2.3_{-2.5}^{+0.7}$ & $0.032_{-0.025}^{+0.012}$ &
$5911.5/5653$\tablenotemark{f}
\enddata
\tablecomments{
\tablenotetext{a}{Assuming a distance of $5\,{\rm kpc}$.}
\tablenotetext{b}{Absorbed fluxes in the $0.5$--$10\,{\rm keV}$ energy range.}
\tablenotetext{c}{Only PN data were used.}
\tablenotetext{d}{Joint-fit results of both PN and MOS data.}
\tablenotetext{e}{Joint-fit results of \emph{Chandra} and \emph{Suzaku} data. The date is the weighted-averaged epoch.}
\tablenotetext{f}{$N_{\rm H}$ is linked in the fit for all observations.}
\tablenotetext{g}{Fixed at $\Gamma=1.33$ from the results
of \citet{ern10}.}}
\end{deluxetable*}
\end{turnpage}
All spectral analyses were performed in the \texttt{Sherpa}
environment\footnote{\url{http://cxc.harvard.edu/sherpa/}}. We
tried an absorbed blackbody plus
power-law (BB+PL) model as in previous studies \citep{rit09,gwk10,cpr14}. We used
the interstellar
absorption model \texttt{tbabs} and the solar abundances
were set to
\texttt{wilm} \citep{wam00}.
The \emph{XMM-Newton} spectra from the same epoch were fit with a
single set of parameters.
We found that the \emph{Chandra} and \emph{Suzaku} spectra share similar
best-fit parameters,
suggesting the quiescent property. In order to boost the
signal-to-noise ratio,
we fit them together with the same parameters.\\
\begin{figure}[ht!]
\centering
\includegraphics[width=0.49\textwidth]{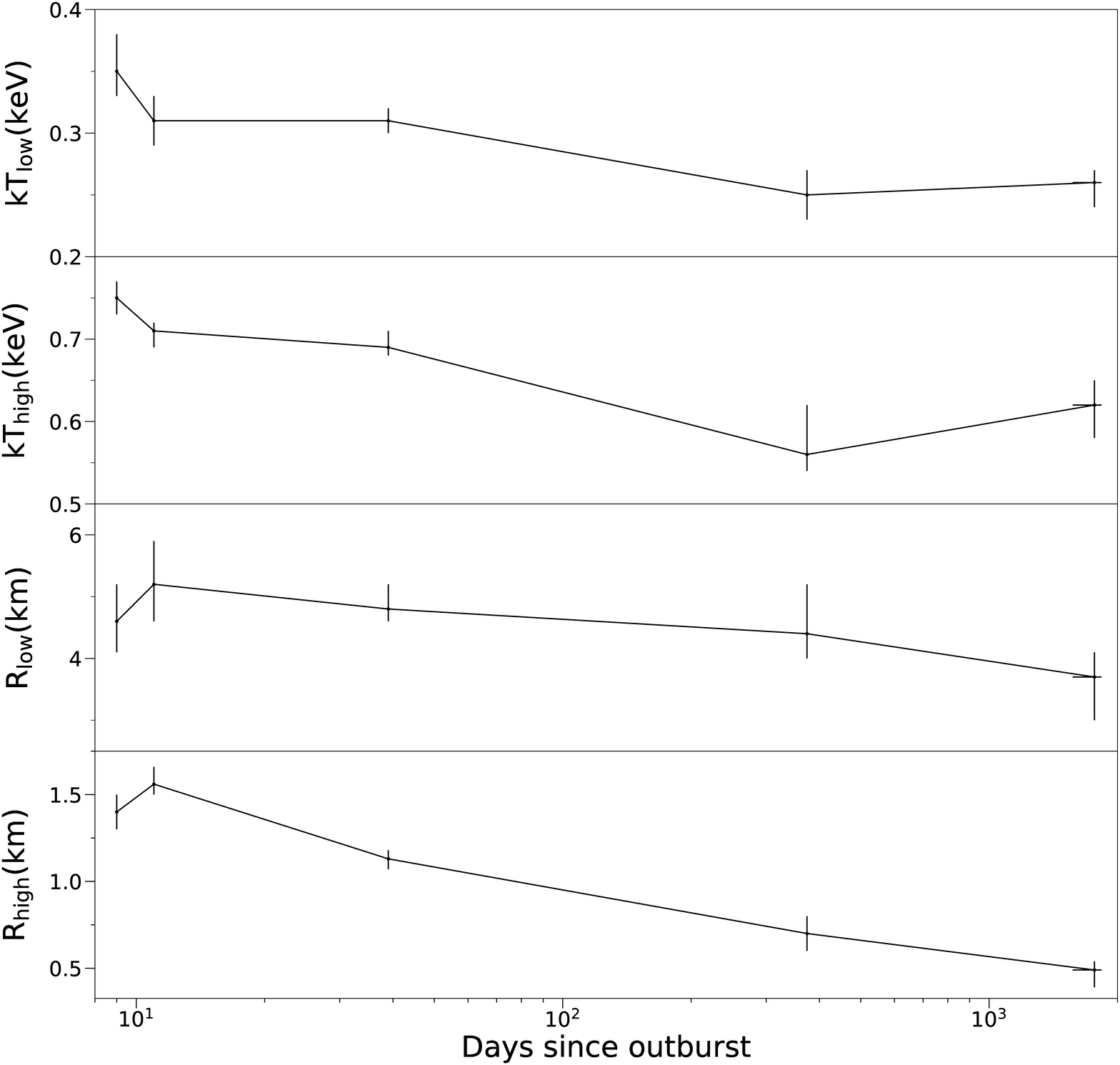}
\caption{Evolution trends for the best-fit parameters of SGR 0501+4516 with
the 2BB+PL model since the 2008 outburst.}\label{fig:trends}
\end{figure}
\indent
The best-fit spectral
parameters are listed in Table\,\ref{tab:pas_results}.
From 2008 to 2013, the best-fit blackbody radius shrunk
significantly from $R=1.45\,{\rm km}$ to $R=0.34\,{\rm km}$ (assuming
$d=5\,{\rm kpc}$) and the power-law photon index
softened from $\Gamma=2.9$ to $\Gamma=3.9$.
Our \emph{XMM-Newton} results are consistent with those reported by
\citet{rit09} and \citet{cpr14} except with a slightly higher absorption column
density $N_{\rm H}$ due to the different absorption model we used. While \citet{cpr14}
suggested that the source had already returned
to its quiescence one year after the 2008 outburst, our new results show that
the total absorbed flux was still
decreasing from $2.8\times10^{-12}\,{\rm erg}\,{\rm cm}^{-2}\,{\rm s}^{-1}$
in 2009 to $2.0\times10^{-12}\,{\rm erg}\,{\rm cm}^{-2}\,{\rm s}^{-1}$ in
2013.
Comparing with the previously reported \emph{Suzaku} results,
our blackbody component has a higher temperature and smaller size. This could
be the
result of the much lower column density ($N_{\rm H}\sim0.4\times10^{22}\,{\rm
cm}^{-2}$) reported by \citet{esk17}. \\
\indent We noted the best-fit PL component is
soft with $\Gamma\gtrsim3$. This could indicate the thermal nature of the
emission. To verify that, we tried to narrow down the energy range to $6\,{\rm keV}$ and
compared the best-fit results of the BB+PL and the double-blackbody (2BB) models. We
found that the latter provided better fits to all spectra, thus, confirming
our idea. When we fit the entire energy range, the 2BB fit has obvious
residuals in the highest energy bins for all \emph{XMM-Newton} spectra, hinting at an additional PL component.
In the final model, we consider the double-blackbody plus power-law (2BB+PL)
model and found that it provides the best fit.
Assuming that $N_{\rm H}$ remained unchanged between epochs, we fit
all spectra simultaneously with a linked absorption model.
We found that the PL component dominated only above $\sim6\,{\rm keV}$ for which our
observations were not very sensitive. As the first three
\emph{XMM-Newton} observations were taken within $\sim1$ month after the
2008\,August\,26 \emph{Suzaku} observation, we believe that they should share
a similar spectral property. In order to obtain a better
fit, we adopted the 2BB+PL result reported by
\citet{ern10} and fixed $\Gamma=1.33$ in the fitting of the 2008
\emph{XMM-Newton} spectra. As the photon index could have changed after 2008, we
did not fix $\Gamma$ for all spectra taken after 2009. However, the PL
component was poorly constrained. We list the best-fit spectral results
in Table\,\ref{tab:pas_results}. The best-fit 2BB+PL model to the \emph{XMM-Newton} PN spectra at different 
epochs are plotted in Figure\,\ref{fig:xmm_fits}, and the fit to the last epoch \emph{Chandra} and \emph{Suzaku} 
spectra are plotted in Figure\,\ref{fig:suzaku_chandra}. Figure\,\ref{fig:trends} shows evolution trends of the
two blackbody components. The
temperature of the cooler blackbody component dropped from
$kT_{\rm low}=0.35\,{\rm keV}$ in 2008 to $0.26\,{\rm keV}$ in 2012,
while there was no significant change in the radius, with $R_{\rm low}$ stayed
$\sim4.5\,{\rm km}$ among all the observations.
The best-fit parameters for the hotter blackbody component, meanwhile,
are consistent with those from the BB+PL fit. Both the temperature
$kT_{\rm high}$ and the radius $R_{\rm high}$ of this component
dropped since the outburst.
We found that adding the best-fit ${\rm BB}_{\rm high}$ component shares
similar parameters as the ${\rm BB}_{\rm high}$ in the BB+PL model.
Similar to the BB+PL results, Figure\,\ref{fig:trends} shows that
$R_{\rm high}$ was not lowest in 2009, indicating that the source was
not yet in quiescence at that time.\\
\begin{figure}[t!]
\centering
\includegraphics[width=0.49\textwidth]{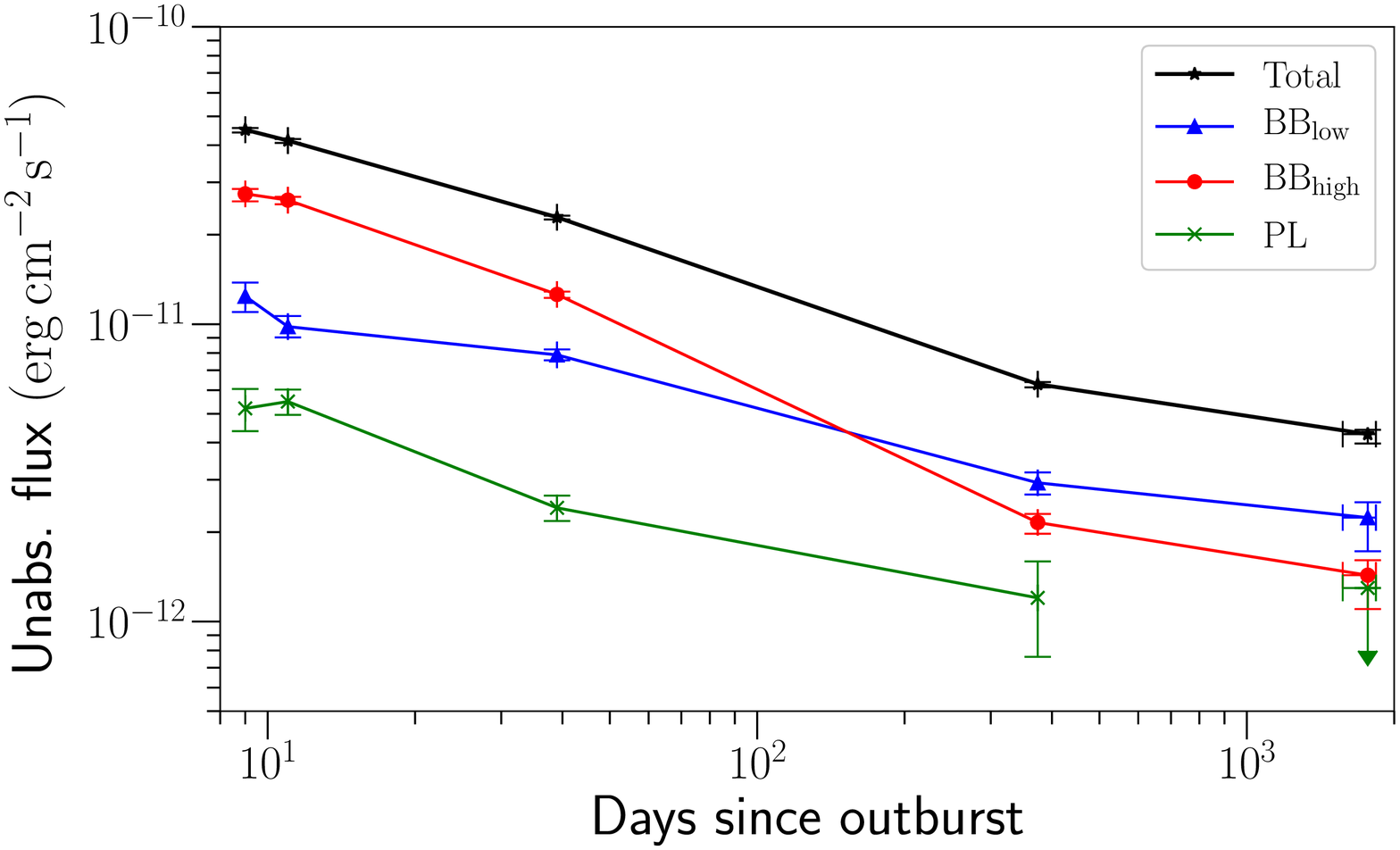}
\caption{Decay trends of unabsorbed fluxes of SGR 0501+4516 for all components
in the 2BB+PL model.}\label{fig:flux}
\end{figure}
\indent In Figure\,\ref{fig:flux}, we plot the flux evolution of all
components in the 2BB+PL
model, we see decreasing trends since the 2008 outburst.
The plot indicates a significant drop of the ${\rm BB}_{\rm high}$ flux after
2009,
and we claim that the source had not yet returned to quiescence at that time.
On the other hand, we found similar
count rates in the 2012 and 2013 \emph{Chandra} observations, which suggests
that
SGR 0501+4516 had reached a quiescent state five years after the outburst.
Finally, we note that there is no obvious plateau in the flux evolution,
contrary to what the crustal cooling model suggests
\citep{let02}.\\
\indent In addition to the BB+PL and 2BB+PL models,
we also tried the resonant cyclotron scattering (RCS) \citep{rzt08} and the 3D
surface thermal emission
and magnetospheric scattering (STEM3D) models \citep{wg15} but
the fits converged to the boundary of the parameter
space. Therefore, we do not believe that the results are physical.\\
\begin{turnpage}
\begin{deluxetable*}{lccccccccc}
\centering
\tablecaption{Two-Temperature Fits to the Spectra of Magnetars in Quiescence
with Uncertainties or Upper Limits at the $90\%$ Confidence Level
\label{tab:low_nh}}
\tablewidth{0pt}
\tabletypesize{\scriptsize}
\tablehead{ \colhead{Object} & Instrument (ObsID) & Model & \colhead{$N_{\rm H}$} &
\colhead{$kT_{\rm low}$} & \colhead{$R_{\rm low}$} & \colhead{$kT_{\rm high}$}
& \colhead{$R_{\rm high}$} & \colhead{$\Gamma$} & $\chi^2_\nu$\,(dof)
\\& & & \colhead{$(10^{22}\,{\rm cm}^{-2})$} & \colhead{(keV)} & \colhead{(km)}
& \colhead{(keV)} & \colhead{(km)} & &
}
\startdata
CXOU J010043.1--721134 &\emph{XMM}\,(see Reference 1) & 2BB & $0.063^{+0.020}_{-0.016}$ & $0.30\pm0.02$ &
$12.1^{+2.1}_{-1.4}$ & $0.68^{+0.09}_{-0.07}$ & $1.7^{+0.6}_{-0.5}$ & \nodata &
$1.14\,(100)$ \\
SGR 0526--66 & \emph{CXO}\,(10806) & 2BB\tablenotemark{a} & $0.07_{-0.05}^{+0.06}$ &
$0.42_{-0.05}^{+0.04}$ & $8.8_{-1.5}^{+1.7}$ &
$1.1_{-0.3}^{+0.8}$ & $0.8_{-0.5}^{+0.9}$ & \nodata &1.31\,(117) \\
XTE J1810--197\tablenotemark{b}& \emph{XMM}\,(see Reference2) & 2BB & $0.60\pm0.02$
& $0.167\pm0.006$ & $9.3\pm1.1$ &
$0.33\pm0.02$ & $0.9\pm0.2$ & \nodata
& $1.21\,(824)$\tablenotemark{c}\\
\emph{Swift} J1822.3--1606 & \emph{CXO}\,(15989-15993) & 2BB & $0.62\pm0.05$ &
$0.11\pm0.01$ & $6.3\pm1.7$ & $0.29\pm0.03$ & $0.24_{-0.10}^{+0.14}$ & \nodata &
1.06\,(74)\\
4U 0142+61\tablenotemark{b} &\emph{XMM}\,(see Reference 3)& 2BB+PL & $0.70\pm0.03$ &
$0.27\pm0.02$ & $14\pm3$ &
$0.50\pm0.02$ & $2.6\pm1.1$ & $2.6\pm0.2$ & $1.11\,(2350)$\tablenotemark{c}\\
SGR 0501+4516 & see Table\,\ref{tab:pas_results}& 2BB+PL & $0.9\pm0.2$ &
$0.26_{-0.02}^{+0.01}$ & $3.7_{-0.7}^{+0.3}$ & $0.62_{-0.04}^{+0.03}$
& $0.49_{-0.10}^{+0.05}$ & $2.3_{-2.5}^{+0.7}$ &
$1.05\,(5653)$\tablenotemark{c} \\
1E 2259+586 & \emph{XMM}\,(0203550701) & 2BB+PL & $1.1\pm0.2$ & $0.32^{+0.04}_{-0.05}$ &
$5.6\pm1.6$ & $0.5^{+0.1}_{-0.2}$ & $1.5^{+1.7}_{-0.7}$ & $3.0^{+0.5}_{-3.2}$
& 1.03\,(494)\\
1E 1048.1--5937& \emph{XMM}\,(0723330101) &2BB+PL & $1.6^{+0.2}_{-0.6}$ &
$<0.18$ & $<410$ &
$0.62\pm0.01$ & $1.7\pm0.5$ & $3.2\pm0.1$ & 0.97\,(909)\\
1RXS J170849.0--400910 & \emph{CXO}\,(4605) & 2BB+PL & $2.72^{+0.04}_{-0.67}$
& $<0.14$ &
$<450$ & $0.41\pm0.05$ & $2.8^{+2.2}_{-1.1}$ & $3.10^{+0.03}_{-0.64}$ &
1.15\,(389)\\
1E 1547.0--5408 & \emph{XMM}\,(0402910101) &2BB\tablenotemark{a} & $3.8^{+0.8}_{-0.6}$ &
$0.39_{-0.08}^{+0.07}$ & $0.9_{-0.4}^{+0.9}$ &
$0.8_{-0.1}^{+0.3}$ & $0.11_{-0.07}^{+0.12}$ & \nodata & 1.52\,(85)\\
SGR 1900+14 & \emph{XMM}\,(0506430101) & 2BB+PL & $4.1^{+0.7}_{-0.2}$ & $<0.12$ & $<1080$ &
$0.39^{+0.01}_{-0.05}$ & $5.5\pm1.9$ & $2.2^{+0.3}_{-0.1}$ & 1.01\,(276)\\
1E 1841--045 & \emph{XMM}\,(0013340101) & 2BB+PL & $4.2^{+1.8}_{-1.1}$ & $<0.28$ & $<2300$ &
$0.5^{+0.1}_{-0.2}$ & $6^{+30}_{-4}$ & $1.9^{+0.4}_{-0.7}$ & 1.12\,(232)\\
CXOU J171405.7--381031 & \emph{CXO}\,(11233) & 2BB+PL & $6.6^{+1.1}_{-1.5}$ &
$<0.28$ & $<90$
& $0.5\pm0.1$ & $<4.3$ & $1.3^{+1.8}_{-2.1}$ & 1.08\,(108)\\
CXOU J164710.2--455216 & \emph{CXO}\,(14360) & 2BB+PL & $6.9^{+1.9}_{-1.7}$ &
$<0.18$ &
$<3800$ & $0.47^{+0.19}_{-0.12}$ & $<9$ & $3.2^{+0.6}_{-1.1}$ & 0.91\,(107) \\
SGR 1806--20 & \emph{CXO}\,(7612) & 2BB\tablenotemark{a} & $9.8^{+1.0}_{-0.9}$ &
$0.67_{-0.09}^{+0.11}$ & $1.8_{-0.5}^{+0.8}$ &
$2.1^{+0.7}_{-0.3}$ & $0.22\pm0.08$ & \nodata &
1.15\,(268)
\enddata
\tablecomments{
\tablenotetext{a}{We noted that BB+PL could be a better model (see the text).}
\tablenotetext{b}{The uncertainties have been scaled to the
$90\%$ confidence level.}
\tablenotetext{c}{Joint fits with different observations.}}
\tablerefs{\scriptsize (1) \citet{tem08}; (2) \citet{bid09};
(3) \citet{gdk10}
}
\end{deluxetable*}
\end{turnpage}
\section{\bf{DISCUSSION}}
\subsection{Two-Temperature Spectral Model}\label{sec:2bb_fit}
Our study showed that the spectrum of SGR 0501+4516 is best described by
a two-temperature model. This is similar to the cases of some magnetars,
including
CXOU J010043.1--721134, XTE J1810--197, and 4U 0142+61
\citep{tem08,bid09,gdk10}.
The result motivates us to
test this model on a larger sample of the magnetar population. \\
\indent We identified 15 magnetars with X-ray observations
taken a few
years after their outbursts.
The three sources mentioned above have
previously been fit with two-temperature spectral models.
For the rest, we reduced \emph{Chandra} and \emph{XMM-Newton} observations
and extracted their spectra with the same
procedures as for SGR 0501+4516. We tried both 2BB and
2BB+PL models on all sources and report the one with lower reduced $\chi^2$
value ($\chi^2_\nu$).
Table\,\ref{tab:low_nh} lists our results and those
reported in previous studies.
We found that, for most magnetars with
small $N_{\rm H}\lesssim1\times10^{22}\,{\rm cm}^{-2}$, their spectra are
generally well fit by
the two-temperature spectral model.
The higher temperature blackbody component always has a smaller radius
$R\lesssim3\,{\rm km}$ and vice versa. For the sources with large
$N_{\rm H}$, the lower temperature blackbody component
is not well constrained due to heavy absorption by the ISM below $2\,{\rm
keV}$. There are three exceptional cases: SGR 0526--66, 1E
1547.0--5408, SGR
1806--20, for which $kT_{\rm high}$ seems too high to be physical. We compared the
$\chi^2$-statistics between the 2BB and the BB+PL fits and found that
they are similar. It is therefore possible that the BB+PL model
provides a more physical description of their spectra.\\
\indent Our results hint that the two-temperature components could be a
common feature among magnetars, although not all could be detected due to
interstellar absorption. The physical interpretation
of the two blackbody components will be discussed below.
\begin{figure}[t!]
\centering
\includegraphics[width=0.5\textwidth]{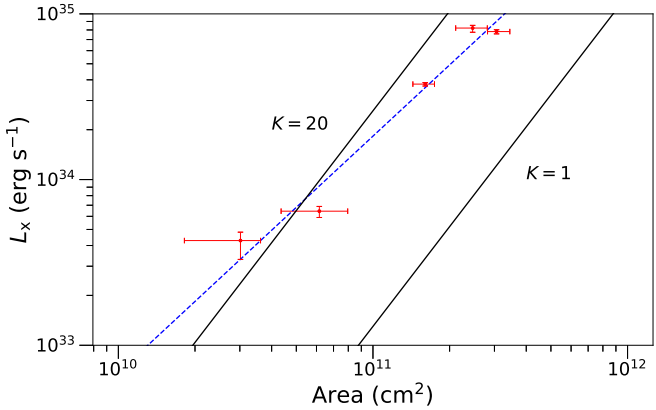}
\caption{Trend of the hotspot X-ray luminosity $L_{\rm X}$ in
$0.5$--$10\,{\rm keV}$ against the hotspot
area $4\pi R_{\rm high}^2$ for SGR 0501+4516. The blue dashed
line shows the best-fit correlation
$L_{\rm X}=1.8\times10^{34}A_{11}^{1.42}\,{\rm erg}\,{\rm s}^{-1}$
and
the black solid lines show the theoretical predicted correlation,
$L_{\rm X}=1.3\times10^{33}KA_{11}^2\,{\rm erg}\,{\rm s}^{-1}$, with $K=1$ and
$K=20$.}\label{fig:l_to_a}
\end{figure}
\begin{deluxetable*}{lccc}[b!]
\tablecaption{Blackbody Temperature and Spin-Inferred Magnetic Field Strength
of Magnetars
and Young High Magnetic Field Rotation-Powered Pulsars as Plotted in
Figure\,\ref{fig:kt_to_b}\label{tab:kt_to_b}}
\tablewidth{0pt}
\tabletypesize{\small}
\tablehead{\colhead{Source} & \colhead{$B$\tablenotemark{a} ($10^{14}\,$G)} &
\colhead{$kT$\tablenotemark{b} (keV)} &
\colhead{Reference}
}
\startdata
\sidehead{Magnetars (entire surface):}
\emph{Swift} J1822--1606 & $0.14$ & $0.11\pm0.01$ & See
Table\,\ref{tab:low_nh}\\
1E 2259+586 & $0.59$ & $0.32_{-0.05}^{+0.04}$ & See Table\,\ref{tab:low_nh}\\
4U 0142+61 & $1.3$ & $0.27\pm0.02$ & 1\\
SGR 0501+4516 & $1.9$ & $0.26_{-0.02}^{+0.01}$ & See Table\,\ref{tab:low_nh}\\
XTE J1810--197 & $2.1$ & $0.167\pm0.006$  & 2\\
CXOU J010043.1--721134 & $3.9$ & $0.30\pm0.02$  & 3\\
1RXS J170849.0--400910 & $4.7$ & $0.42\pm0.02$ & 4\\
SGR 0526--66 & $5.6$ & $0.44\pm0.02$  & 5\\
SGR 1900+14 & $7.0$ & $0.47\pm0.02$ & 6\\
1E 1841--045 & $7.0$ & $0.45\pm0.03$ & 7\\
SGR 1806--20 & $11.3$ & $0.55\pm0.07$ & 8\\
\sidehead{Magnetars (hotspot):}
SGR 0418+5729 & $0.061$ & $0.32\pm0.05$  & 9\\
\emph{Swift} J1822--1606 & $0.14$ & $0.29\pm0.03$ & See
Table\,\ref{tab:low_nh}\\
1E 2259+586 & $0.59$ & $0.5_{-0.2}^{+0.1}$ & See Table\,\ref{tab:low_nh}\\
CXOU J164710.2--455216 & $1.0$ & $0.53\pm0.03$ & 10\\
4U 0142+61 & $1.3$ & $0.50\pm0.02$ & 1\\
SGR 0501+4516 & $1.9$ & $0.62_{-0.04}^{+0.03}$  & See
Table\,\ref{tab:low_nh}\\
XTE J1810--197 & $2.1$ & $0.33\pm0.02$  & 2\\
SGR 1935+2154 & $2.2$ & $0.47\pm0.03$ & 11\\
1E 1547.0--5408 & $2.2$ & $0.43\pm0.05$ & 12\\
PSR J1622--4950 & $2.7$ & $0.5\pm0.1$  & 13\\
CXOU J010043.1--721134 & $3.9$ & $0.68_{-0.07}^{+0.09}$ & 3\\
1E 1048.1--5937 & $4.5$ & $0.56\pm0.02$ & 14\\
\sidehead{High-$B$ rotation-powered pulsars:}
PSR B1509--58 & $0.15$ & $0.15\pm0.01$ & 15\\
PSR J1119--6127 & $0.41$ & $0.21\pm0.04$ & 16\\
PSR J1846--0258 & $0.49$ & $<0.25$ & 17
\enddata
\tablecomments{
\tablenotetext{a}{\scriptsize Adopted from the Magnetar Catalog
\citep{ok14}.}
\tablenotetext{b}{\scriptsize Uncertainties are at the $90\%$ confidence level.}}
\tablerefs{\scriptsize
(1) \citet{gdk10}; (2) \citet{bid09}; (3) \citet{tem08};
(4) \citet{cri07}; (5) \citet{phs12}; (6) \citet{met06b};
(7) \citet{ks10};
(8) \citet{emt07b}; (9) \citet{rip13}; (10) \citet{akac13};
(11) \citet{ier16}; (12) \citet{bis11}; (13) \citet{ags12};
(14) \citet{tgd08}; (15) \citet{hnt17};
(16) \citet{nkh12}; (17) \citet{ln11}
}
\end{deluxetable*}
\subsection{Physical Interpretation of the Hotter Blackbody
Component}\label{sec:high_bb}
The best-fit radius of the higher temperature component $R_{\rm high}$ of SGR~0501+4516
had shrunk to $0.49\,{\rm km}$ from 2008 to 2013,
indicating that the thermal emission could come from a hotspot on surface.
There were several magnetars with
blackbody radii that continued to shrink for a few years after their outbursts
\citep{bl16}.
\citet{bel09} suggested that this
could be the observational evidence supporting the $j$-bundle model. When a
twisted magnetic field is implanted into the closed
magnetosphere, the current ($j$-bundle) would flow along the closed magnetic
field lines and return back to the stellar surface,
heating up the footprints of the $j$-bundle and resulting in hotspots. After
an outburst, the footprints
are expected to keep shrinking and the hotspot could be observed as a blackbody
component with a decreasing radius.
This predicts a correlation between the X-ray luminosity and the area of a
hotspot as
$L_{\rm X}=1.3\times10^{33}KA_{11}^{2}\,{\rm erg}\,{\rm s}^{-1}$, where $A_{11}$ is
the blackbody
area in units of $10^{11}\,{\rm cm}^{2}$ and $K$ is a constant depending on
the twisting angle of the $j$-bundle,
the surface magnetic field strength, and the discharge voltage
\citep{bel09,bel11}.\\
\indent We plot in Figure\,\ref{fig:l_to_a} the hotspot
luminosity of SGR 0501+4516 against
its area $A=4\pi R_{\rm high}^2$. The distance of the source is assumed to be
$d=5\,{\rm kpc}$
for calculating the luminosity. Our result broadly agrees with the theory
prediction and suggests $K\sim20$.
If we fit the data points with a straight line in the log--log plot in
Figure\,\ref{fig:l_to_a},
the best-fit correlation is flatter, with
$L_{\rm X}=1.8\times10^{34}A_{11}^{1.42}\,{\rm erg}\,{\rm s}^{-1}$.
Similar behavior was also found in several other magnetars during flux
relaxation after outbursts \citep{bl16}.
The discrepancy could be due to the time variation of the proportionality
constant $K$.\\
\subsection{Physical Interpretation of the Cooler Blackbody
Component}\label{sec:low_bb}
\begin{figure}[t!]
\centering
\includegraphics[width=0.43\textwidth]{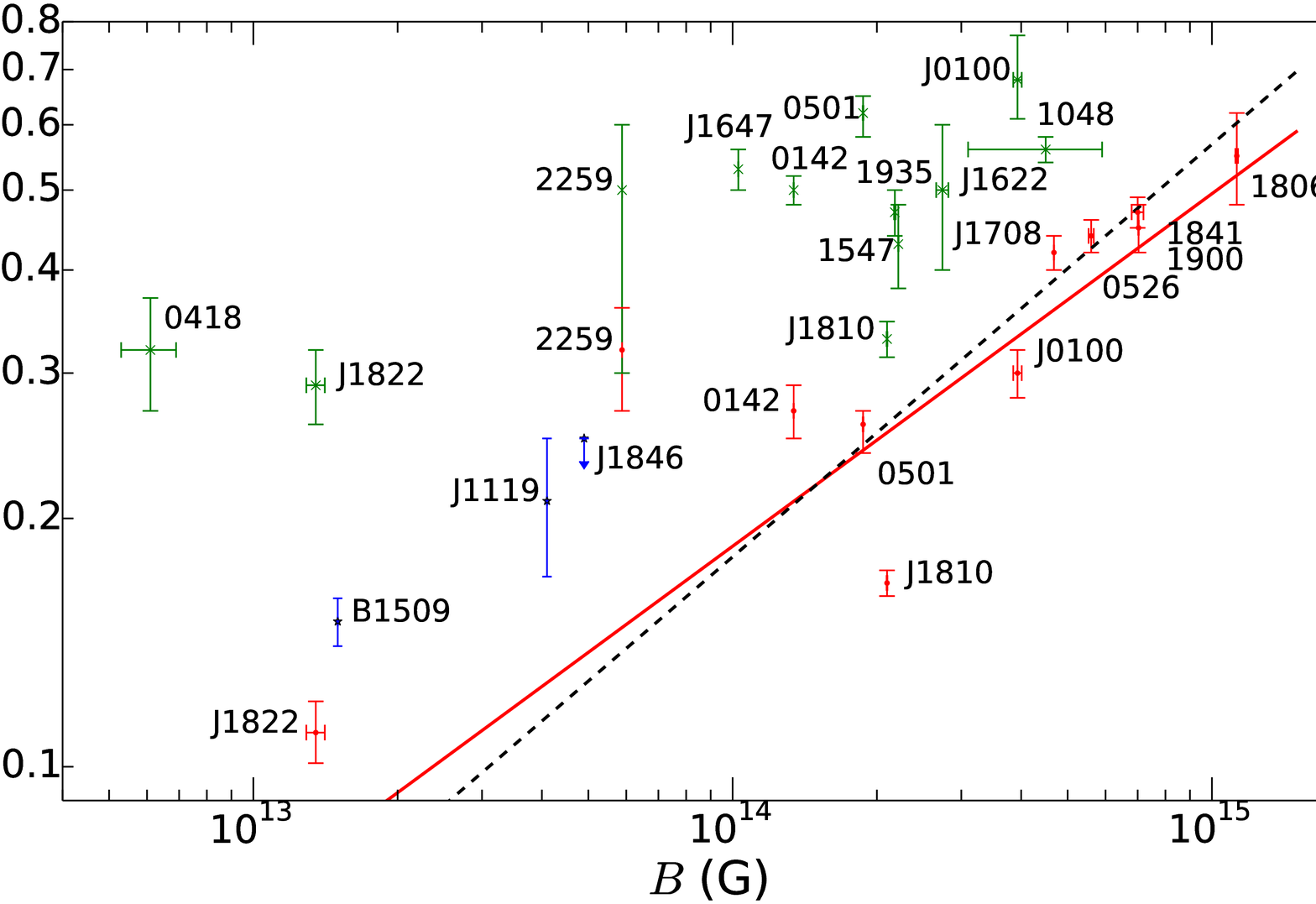}
\caption{Blackbody temperature against magnetic field strength of magnetars
and three young high-$B$ rotation-powered pulsars,
using values listed in Table\,\ref{tab:kt_to_b}.
The red and green dots indicate blackbodies from the entire surface and the
hotspots, respectively.
The blue dots represents the high-$B$ rotation-powered pulsars. The red solid line shows
the best-fit correlation
$kT\propto B^{0.4}$ using the red data points only. The black dashed line
represents the theoretical prediction of $kT\propto
B^{0.5}$.}\label{fig:kt_to_b}
\end{figure}
In the two temperature fits, the cooler blackbody always shows a larger radius
$R_{\rm low}$ and some values listed in Table\,\ref{tab:low_nh} are compatible
with the neutron star
radius. We therefore believe that this blackbody component could originate
from the entire surface. An additional support is that in our case of SGR
0501+4516, $R_{\rm low}$ has been relatively stable during flux relaxation.
Theories suggest that the thermal emission of magnetars
could be arise from the decay of the crustal magnetic field
\citep{td96,plm07}. If this
is the only energy source, one expects
a correlation between the surface temperature $kT$ and the magnetic
field strength $B$ \citep{plm07}. The conservation of energy could be
expressed as
\begin{equation}
-A\Delta R\frac{{\rm d}E_{\rm m}}{{\rm d}t}=A\sigma T^4, \label{eq:b_to_t}
\end{equation}
where $E_{\rm m}$ is the magnetic energy density, $A$ is the emission area,
$\Delta R$ is the thickness of
the neutron star crust, and $\sigma$ is the Stefan-Boltzmann constant. The
magnetic energy density
$E_{\rm m}$ could be written as $B^2/8\pi$. If the decay of $B$ is in the
exponential form, it implies a relation $T\propto \sqrt{B}$.
Note that this ignores any age effects that are justified,
as magnetars are young objects in general \citep[see][]{vrp13}.\\
\indent To verify the correlation above,
we investigated the trend between $kT$ and $B$ for
all quiescent magnetars, using the latest results reported in the literature and
from our own analysis (see Table\,\ref{tab:low_nh}). These values are listed
in Table\,\ref{tab:kt_to_b}
and plotted in Figure\,\ref{fig:kt_to_b}.
As we mentioned, some blackbody components correspond to the hotspot and some
correspond to the entire
surface. We show them separately in the plot as two groups, depending on
whether the blackbody radius $R$
is larger or smaller than $3\,{\rm km}$. The plot shows an increasing trend for the entire surface $kT$,
with a correlation coefficient $r=0.85$. We fit the log--log plot with a
straight line and obtained
$kT\propto B^{0.4}$, which is a bit flatter than, but generally comparable
with the theoretical prediction of $B^{0.5}$.
On the other hand, the temperature of the hotspots
shows no such a correlation, which suggests that they could probably be powered
by $j$-bundle
instead of the decay of the crustal field.\\
\begin{deluxetable*}{lcccc}
\tablecaption{Quiescent X-Ray Luminosity in $2$--$10\,{\rm keV}$ and
Spin-Inferred Magnetic Field Strength
of Magnetars and High Magnetic Field Rotation-Powered Pulsars as Plotted in
Figure\,\ref{fig:l_to_b}\label{tab:l_to_b}}
\tablewidth{0pt}
\tabletypesize{\small}
\tablehead{\colhead{Source} & \colhead{$L_{\rm X}$\tablenotemark{a} ($10^{35}\,{\rm
erg}\,{\rm s}^{-1}$)} &
\colhead{$B$\tablenotemark{b} $(10^{14}\,{\rm G})$} &
\colhead{Distance\tablenotemark{b} (kpc)} & \colhead{Reference}
}
\startdata
\sidehead{Magnetars:}
SGR 0418+5729 & $1.0_{-0.9}^{+1.1}\times10^{-5}$ & $0.061$ &
$2.0\pm0.4$\tablenotemark{c} & 1\\
\emph{Swift} J1822.3--1606 & $5_{-4}^{+3}\times10^{-5}$ & $0.14$ &
$1.6\pm0.3$ & See Table\,\ref{tab:low_nh} \\
1E 2259+586 & $0.20_{-0.06}^{+0.04}$ & $0.59$ & $3.2\pm0.2$ & See
Table\,\ref{tab:low_nh} \\
CXOU J164710.2--455216 & ($4.5\pm3.8)\times10^{-3}$ & $1.0$ & $3.9\pm0.7$ &
2\\
4U 0142+61 & $1.05\pm0.33$ & $1.3$ & $3.6\pm0.4$ & 3\\
SGR 0501+4516 & $3.5_{-1.3}^{+1.0}\times10^{-2}$ & $1.9$ &
$5.0\pm1.0$\tablenotemark{c} & See Table\,\ref{tab:low_nh}\\
XTE J1810--197 & $1.3_{-0.9}^{+0.5}\times10^{-3}$ & $2.1$ & $3.1\pm0.5$ & 4\\
1E 1547.0--5408 & $1.3_{-0.9}^{+0.5}\times10^{-2}$ & $2.2$ & $4.5\pm0.5$ & 5\\
SGR 1627--41 & $2.5_{-1.3}^{+2.3}\times10^{-2}$ & $2.2$ & $11.0\pm0.3$ & 6\\
PSR J1622--4950 & $4.4_{-3.6}^{+7.0}\times10^{-3}$ & $2.7$ &
$9.0\pm1.8$\tablenotemark{c} & 7 \\
CXOU J010043.1--721134 & $0.7_{-0.3}^{+1.7}$ & $3.9$ & $62.4\pm1.6$ & 8\\
1E 1048.1--5937 & $0.5\pm0.3$ & $4.5$ & $9.0\pm1.7$ & 9\\
1RXS J170849.0--400910 & $0.42\pm0.11$ & $4.7$ & $3.8\pm0.5$ & 10\\
CXOU J171405.7--381031 & $0.33\pm0.24$ & $5.0$ & $10.2\pm3.5$ & 11\\
SGR 0526--66 & $1.9_{-0.4}^{+0.3}$ & $5.6$ & $53.6\pm1.2$ & 12\\
SGR 1900+14 & $0.7\pm0.3$ & $7.0$ & $12.5\pm1.7$ & 13\\
1E 1841--045 & $1.8_{-1.0}^{+0.7}$ & $7.0$ & $8.5_{-1.0}^{+1.3}$ & 14\\
SGR 1806--20 & $1.6_{-0.7}^{+0.8}$ & $11.3$ & $8.7_{-1.5}^{+1.8}$ & 15\\
\sidehead{High-$B$ rotation-powered pulsars:}
PSR B1509--58 & $0.96\pm0.05$ & $0.15$ & $5.2\pm1.4$ & 16 \\
PSR J1119--6127 & $2.5_{-1.3}^{+3.2}\times10^{-3}$ & $0.41$ & $8.4\pm0.4$ & 17
\\
PSR J1846--0258 & $0.19_{-0.03}^{+0.04}$ & $0.49$ & $6.0_{-0.9}^{+1.5}$ & 18
\enddata
\tablecomments{
\tablenotetext{a}{\scriptsize $90\%$ uncertainties in
$L_{\rm X}$, derived by
combining the errors in flux
and distance using the standard error propagation formula.}
\tablenotetext{b}{\scriptsize Adopted from the Magnetar Catalog
\citep{ok14}. For those with multiple estimated distances, we
simply used the most updated or the better measured values.}
\tablenotetext{c}{\scriptsize As the uncertainty in distance is not
reported, we assumed a relative error of $20\%$, similar to that of other
sources.}}
\tablerefs{\scriptsize
(1) \citet{rip13}; (2) \citet{akac13};
(3) \citet{rni07}; (4) \citet{bid09}; (5) \citet{gg07};
(6) \citet{eiz08}; (7) \citet{ags12}; (8) \citet{tem08};
(9) \citet{tgd08};
(10) \citet{rio07}; (11) \citet{sbni10}; (12) \citet{phs12};
(13) \citet{nmy09}; (14) \citet{ks10}; (15) \citet{emt07b};
(16) \citet{hnt17}; (17) \citet{nkh12}; (18) \citet{ln11}
}
\end{deluxetable*}
\indent There is recent evidence showing that both
young high magnetic field rotation-powered pulsars and magnetars share
similar properties, making the division between these two classes blurred
\citep[see][]{ggg08,nk11,glk16}. This motivates us to include the three young
sources with age of $\sim10^3\,{\rm years}\,$, PSRs B1509--58, J1119--6127, and J1846--0258,
in Table\,\ref{tab:kt_to_b} and Figure\,\ref{fig:kt_to_b} for comparison.
The thermal emission of PSRs B1509--58 and J1119--6127 has blackbody radii
$R=10^{+39}_{-5}\,{\rm km}$
and $3^{+4}_{-1}\,{\rm km}$, respectively, suggesting that they could be
originated from
the entire surface (or large area; 
\citealt{nkh12,hnt17}). However, for PSR B1509--58, the blackbody radius was not
very well
constrained due to strong non-thermal emission.
On the other hand, there is no thermal emission found in PSR J1846--0258 in
quiescence, with an
upper limit of $0.25\,{\rm keV}$ \citep{ln11}. From the plot,
it is interesting to note that all high-$B$ rotation-powered pulsars seem to follow the
same $kT$--$B$ trend
as magnetars. Our results suggest that the energy source, i.e. $B$-field
decay, could power the
entire surface thermal emission of magnetars and high-$B$ rotation-powered pulsars.\\
\indent
While $kT$ and $B$ appear to show a correlation that is broadly consistent
with the theory,
there remain some unsolved problems in this picture.
The temperature of the cooler blackbody component is typically higher in
outburst, then decays to a constant value a few years after.
Hence, the outburst could partly contributed to
the thermal emission (see Figure\,\ref{fig:trends} and also
\citealt{bid09} and \citealt{gdk10}). Also, we note that
some radii of the cooler blackbody are
smaller than that of a neutron star.
It could indicate that the emission regions are smaller than the entire
surface or that the temperature
distribution is inhomogeneous. It is unclear if the Equation\,\ref{eq:b_to_t}
needs to be modified in this case.
\begin{figure}[ht!]
\centering
\includegraphics[width=0.41\textwidth]{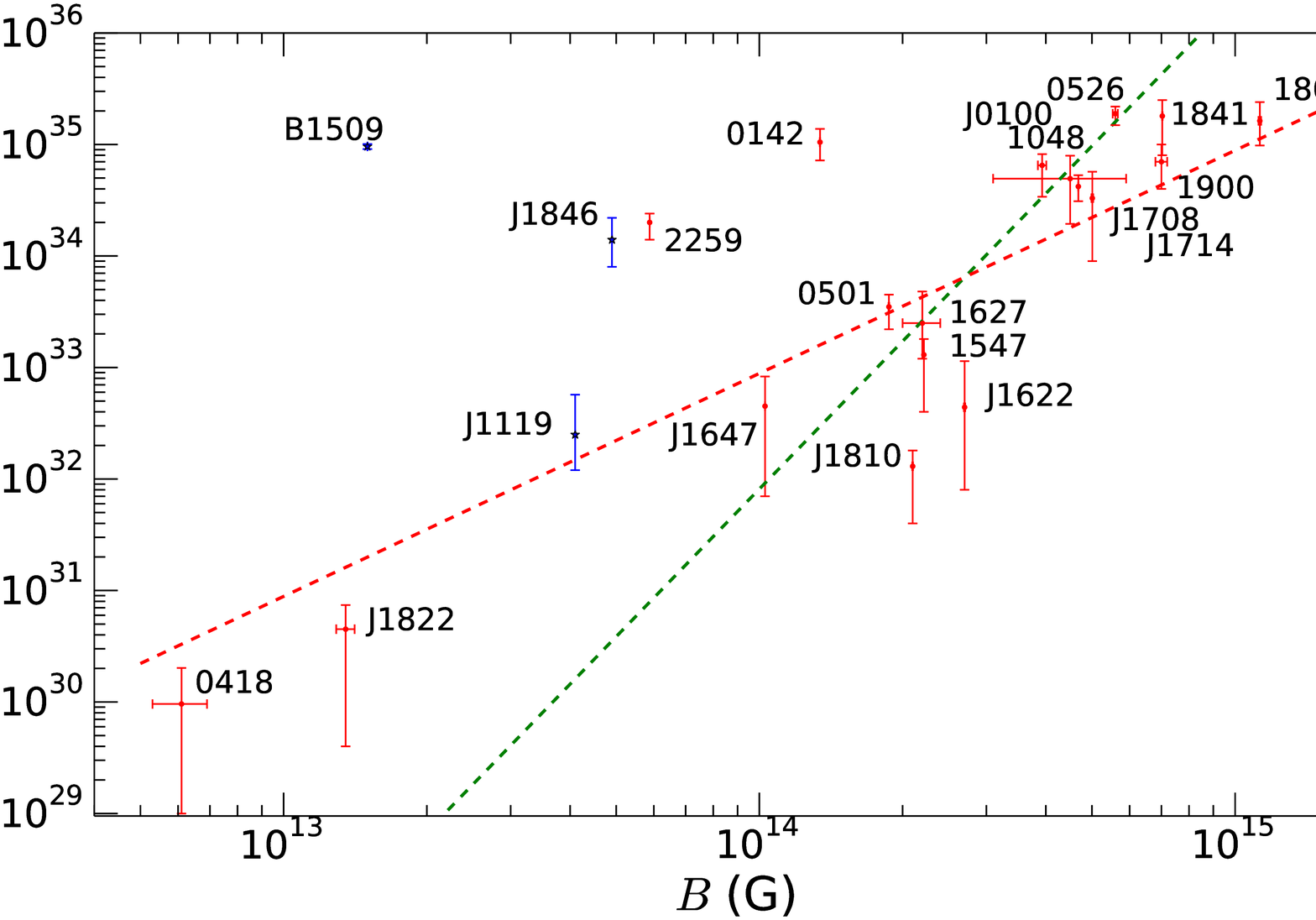}
\caption{X-ray luminosity $L_{\rm X}$ in $2$--$10\,{\rm keV}$ against
magnetic field strength $B$, using values listed in Table\,\ref{tab:l_to_b}. The red dots indicate magnetars in quiescence, and
the blue dots indicate high-$B$ rotation-powered pulsars.
The red and the green dashed lines show the theoretical predictions of
$L_{\rm X}\propto B^{2}$ and $\propto B^{4.4}$,
respectively.}\label{fig:l_to_b}
\end{figure}
\subsection{Correlation Between X-Ray Luminosity and Magnetic
Field}\label{sec:correlation_L_kT_B}
We revisit the correlation between the quiescent X-ray luminosity, $L_{\rm X}$, and
the magnetic field, $B$, of magnetars as reported by \citet{akt12}, using
updated measurements
listed in Table\,\ref{tab:l_to_b}. The results are plotted in
Figure\,\ref{fig:l_to_b}.
We compared the trend with two theoretical predictions of $L_{\rm X}\propto
B^{2}$ deduced from the Equation\,\ref{eq:b_to_t} \citep{plm07} and
$L_{\rm X}\propto B^{4.4}$ based on the ambipolar diffusion model with neutrino
cooling \citep{td96}.
The plot shows a general trend but with large scatter, particularly for
magnetars
with $B\sim10^{14}\,{\rm G}$.
Our updated plot prefers $B^2$, providing some support to the
simple magnetic field decay model.
Note that our result contradicts that reported by \citet{akt12}.
The main discrepancy is due to the updated measurements from two low-field
magnetars, SGR 0418+5729 and
\emph{Swift} J1822.3--1606. If we fit the log--log plot with a straight line,
we obtain a slightly
flatter correlation of $L_{\rm X}\propto B^{1.7}$.
From the plot, 1E~2259+586 and 4U~0142+61 are far more
luminous than other magnetars with similar $B$. Excluding these two
outliers gives $L_{\rm X}\propto B^{2.8}$, which again prefers $B^2$ to $B^{4.4}$.\\
\indent Similar to the $kT$--$B$ plot, we also include three young high
magnetic field rotation-powered pulsars in Figure\,\ref{fig:l_to_b}.
We found that only PSR J1119--6127 follows the general trend of magnetars,
while the other two, PSRs B1509--58 and J1846--0258, have luminosities a few
orders of magnitude
higher. We believe that their X-ray emission is dominated by
non-thermal radiation powered by spin-down, which could be a main difference
between magnetars and high-$B$ rotation-powered pulsars. Although the correlation appears
to support to the
theoretical prediction, there are too few magnetar examples with
$B<10^{14}\,{\rm G}$.
Increasing the sample in this magnetic field range in future studies can
better confirm the theory.
\section{\bf{CONCLUSION}}\label{sec:con}
We performed spectral and timing analyses of SGR 0501+4516 using new and
archival
X-ray observations taken with \emph{Chandra}, \emph{XMM-Newton}, and
\emph{Suzaku}. We show that the
source has returned to quiescence in 2013, five years after the outburst.
Our timing results found a spin period of $\sim5.762\,{\rm s}$  with stable
pulse profiles in 2012 and 2013.
The \emph{Chandra} images show no detectable proper motion, with an upper
limit of $0\farcs32\,{\rm yr}^{-1}$,
rejecting the idea that SGR 0501+4516 was born in SNR G160.9+2.6.
We found that the soft X-ray spectrum is best described by
a double blackbody plus power-law (2BB+PL) model.
The quiescent spectrum has temperatures of $0.26\,{\rm keV}$
(with $R=3.7\,{\rm km}$) and $0.62\,{\rm keV}$ (with $R=0.49\,{\rm km}$).
We found a correlation between the X-ray luminosity and the area of the
evolving hotter blackbody component, which agrees with the prediction of the
$j$-bundle model.\\
\indent We further applied the two-temperature spectral model to other
magnetars in
quiescence and found that it provides a good fit to
most sources with low column density,
suggesting that this could be a common feature.
We investigated the correlation between the blackbody temperature $kT$ and the
spin-inferred
magnetic field $B$ of all magnetars in quiescence.
For blackbodies with large areas comparable to the entire stellar surface,
the correlation generally agrees with the prediction from the simple magnetic
field decay model.
We found that this simple scenario can also explain the trend between the
quiescent X-ray luminosity and magnetic field strength of magnetars.

\acknowledgements
We thank the referee for the comments that improved
this paper. The scientific results reported in this article are based on observations made
by the
\textit{Chandra X-ray Observatory} and data obtained from the \textit{Chandra}
Data Archive.
This work was based on observations obtained with \textit{XMM-Newton}, an ESA
science mission with instruments and contributions directly funded by ESA
Member States and NASA.
This research has made use of the NASA Astrophysics Data System (ADS) and
software provided by the Chandra X-ray Center (CXC) in the application package
CIAO and Sherpa.
This work is supported by a GRF grant of Hong Kong Government under HKU
17300215P.

\end{document}